\documentclass[12pt]{article}
\usepackage{graphicx}
\usepackage{cite}
\usepackage{amsmath}
\usepackage{amssymb}
\usepackage{enumitem}
\usepackage{latexsym}
\begin{document}
%\begin{flushright}
%BHU-PHYS-CAS Preprint\\
%arXiv: 1005.5067 [hep-th]
%\end{flushright}
\vskip 2cm

\def\warning#1{\begin{center}
\framebox{\parbox{0.8\columnwidth}{\large\bf #1}}
\end{center}}

\begin{center}

{\large {\bf Novel Symmetries in Christ-Lee Model}}

\vskip 2.5 cm

{\sf{ \bf R. Kumar$^{a ,}$\footnote{Present address: Department of Physics \& Astrophysics, University of Delhi, Delhi--110007} and A. Shukla$^b$}}\\
\vskip .1cm
{\it $^a$S. N. Bose National Centre for Basic Sciences,\\ Salt Lake, Kolkata-700098, West Bengal, India}\\
{\it $^b$Indian Institute of Science Education and Research, \\Mohanpur--741246, Kolkata, India}\\
\vskip 0.1 cm
{\sf {E-mails: raviphynuc@gmail.com; as3756@iiserkol.ac.in}}\\
\end{center}

\vskip 2.5 cm

\noindent

\noindent
{\bf Abstract:} We demonstrate that the gauge-fixed Lagrangian of the Christ-Lee model respects {\it four} 
fermionic symmetries, namely; (anti-)BRST symmetries, (anti-)co-BRST symmetries within the framework of BRST formalism. 
The appropriate anticommutators amongst the fermionic symmetries lead to a {\it unique} bosonic symmetry. It turns out that the algebra
obeyed by the symmetry transformations (and their corresponding conserved charges) is reminiscent of the algebra satisfied 
by the de Rham cohomological operators of differential geometry. We also provide the physical realizations of the cohomological
operators in terms of the symmetry properties. Thus, the present model provides a simple model for the Hodge theory.



\newpage

\section{Introduction}

Symmetry principles play an important role to understand the laws of nature. The local gauge symmetry is one of the important symmetries   
in physics that relates the internal space with external spacetime. These symmetry transformations are generated due to the existence of  
the first-class constraints in a given theory \cite{dira, sund}. In order to quantize the system with first-class constraints, the 
Becchi-Rouet-Stora-Tyutin (BRST) formalism plays a decisive role \cite{brs1, brs2, brs3, brs4}. In this formalism, for a local gauge symmetry 
associated with a given classical system,  we have two linearly independent supersymmetric type BRST and anti-BRST symmetries at the quantum level. 
The geometrical interpretation and the origin of nilpotent and absolutely anticommuting (anti-)BRST symmetry transformations have also been shown
within the framework of superfield formalism \cite{bt1, bt2, del}.

The BRST formalism is one of the most institutive and elegant approaches to quantize the
$p$-form ($p = 1, 2, 3,...$) gauge theories as well as it also encompasses many mathematical aspects in its fold. For instance, 
the mathematical aspects of BRST formalism have a deep connection
with the abstract de Rham cohomological operators of differential geometry \cite{rpm1, rpm2, rpm3}. 
In our earlier work \cite{rpm4}, we have established that any arbitrary Abelian $p$-form ($p = 1, 2, 3$) gauge theory  in $D = 2p$ dimensions of spacetime, besides the usual (anti-)BRST symmetries,  also endowed with the (anti-)co-BRST symmetries within the framework of BRST formalism. 
Thus, in $D = 2p$ dimensions of spacetime, we have been able to show that Abelian $p$-form theory provides a field-theoretic model for Hodge theory 
where the de Rham cohomological operators find their physical realizations in terms of the continuous symmetry transformations (and corresponding conserved charges). The discrete symmetry provides the analogue of the Hodge duality ($*$) operation. 
Recently, the rigid rotor as a toy model for the Hodge theory has also been illustrated in a great detail \cite{rpm5}.
In a recent set of papers \cite{rpm6, rpm7, rpm8}, it has been shown that the six continuous symmetries, present in a class of gauge 
field-theoretic models which happen to be models for the Hodge theory, 
lead to the same and unique canonical brackets amongst the creation and annihilation operators without taking any help of the canonical momenta.  
These canonical brackets play an important role in the quantization of the system.

It is worthwhile to mention that the BRST symmetries and related geometric approach have played a key role in 
our understanding of the quantization of superstring, supergravity, supersymmetric gauge theories 
and higher $p$-form ($p \geq 2$) gauge theories. Recently, the nilpotent BRST and anti-BRST symmetries
for the perturbative quantum gravity have been established in (non-)linear gauges \cite{mir1, mir2}. 
Furthermore, the perturbative quantum gravity has also been studied in complex and noncommutative spacetime within the framework of BRST formalism
\cite{mir3,mir4,mir5}. The superspace formulation of higher derivative theories \cite{mir6},  Chern-Simons, Yang-Mills theories in the context of Batalin-Vilkovisky formalism \cite{mir7,mir8} and ABJM theory on deformed superspace \cite{mir9,mir10} have also been analyzed.

The Christ-Lee  model is a classical constrained system which has been well-studied at the quantum level, too \cite{cl}. 
It is a gauge invariant model having 
two independent first-class constraints in the language of  Dirac's classification scheme of constraints. 
As far as the quantization of the Christ-Lee model is concerned, it has been studied in many different perspectives \cite {co1,co2, bl, emr}.
The consistent BRST quantization of this model is also carried out  \cite{kul}.  
The purpose of our  present investigation is two-fold. First, we explore the symmetries associated with the Christ-Lee model.
In the literature \cite{kul}, only the off-shell nilpotent (anti-)BRST symmetries were known. In our present work, we show that, in 
addition to the usual (anti-)BRST symmetries, the nilpotent and absolutely anticommuting (anti-)co-BRST (also known as 
(anti-)dual-BRST) symmetries and a unique bosonic symmetry do exist for the present model. Second, the development of present 
model as a  physical model for the Hodge theory is always interesting because of the fact that the abstract mathematical operators of differential 
geometry get their physical realizations in terms of the continuous symmetry transformations (and corresponding conserved charges).
Thus, the Christ-Lee model provides a simple  model for the Hodge theory within the framework of BRST formalism.

The contents of the present paper are as follows. In next section, we briefly discuss about the Christ-Lee model and associated  local gauge 
symmetry. Third section is devoted for the discussion about the off-shell nilpotent and absolutely anticommuting (anti-)BRST symmetry transformations (and 
their corresponding charges). We show, in the fourth section,  that the off-shell nilpotent and absolutely anticommuting (anti-)co-BRST symmetries 
also exist for the present model.  In the fifth section, we discuss a unique bosonic symmetry and its corresponding generator. The bosonic symmetry 
emerges due to the existence of the non-vanishing anticommutators amongst the above fermionic symmetries. 
The sixth section is devoted for the discussion of  the extended BRST 
algebra satisfied by the symmetry transformations (and corresponding conserved charges) and we also show that this algebra is reminiscent of the Hodge 
algebra obeyed by the de Rham cohomological operators of  differential geometry . In this section, the physical realizations of the cohomological
operators are also captured in terms of the symmetry properties. Finally, in the last section, we provide some concluding remarks and point out  
the future directions regarding our present investigation.

\section{Christ-Lee model: local gauge symmetry}
We begin with the Christ-Lee  model in terms of the plane polar coordinates ($r, \theta$) \cite{cl, co2, kul}. 
The first-order Lagrangian for the Christ-Lee model is given as
\begin{eqnarray}
L_f = \dot r \,p_r + \dot \theta\, p_\theta - \frac{1}{2}\, p^2_r - \frac{1}{2 r^2}\, p^2_\theta - z \,p_\theta - V(r),
\end{eqnarray}
where $\dot r$ and $\dot \theta$ define the generalized velocities, $z$ is another generalized coordinate, $p_r$ and $p_\theta$ 
are the canonically conjugate momenta corresponding to the variables $r$ and $\theta$, respectively.  $V(r)$ is the potential bounded from below 
and all the variables are function of time evolution parameter $t$. It is clear that the velocity $\dot z$ is missing in the Lagrangian. 
As a consequence, the present system has a primary constraint:  
\begin{eqnarray}
\Omega_1 = \frac{\partial L_f}{\partial {\dot z}} = p_z \approx 0, 
\end{eqnarray}
where $p_z$ is the momentum corresponding to the auxiliary variable $z$. 
The symbol $\approx$ defines weak equality in the sense of Dirac.  
The time evolution of the primary constraint $\Omega_1$ leads to the following secondary constraint: 
\begin{eqnarray}
\frac{d \Omega_1}{dt}  = \frac{d}{dt} \Big(\frac{\partial L_f}{\partial {\dot z}}\Big) \approx 0 \Rightarrow  \Omega_2 = p_\theta \approx 0.
\end{eqnarray}
It can be explicitly checked that the time evolution of $\Omega_2$ does not yield any further constraint on the 
theory because $\Omega_2$ commutes with the Hamiltonian of the present model. As a consequence, the Christ-Lee
model endowed only with two first-class constraints in the Dirac's terminology \cite{dira, sund}. 
These first-class constraints are the generator of the gauge symmetry present in the system. 
The gauge symmetry generator can be written as \cite{sund, rothe}  
\begin{eqnarray}
G = {\dot \chi}\,\Omega_1 + \chi\, \Omega_2,
\end{eqnarray}
where $\chi(t)$ is (time-dependent) local gauge parameter. Using the definition of a generator
\begin{eqnarray}
\delta \;\xi(t) = - i\, [ \xi(t),\; G], 
\end{eqnarray}
where $\xi(t)$ denotes any generic variable present in the model, we obtain the following gauge transformations: 
\begin{eqnarray}
\delta\, z = \dot \chi(t), \qquad  \delta \,\theta = \chi(t), \qquad  \delta[r,\, p_r,\, p_\theta, \; V(r)].
\end{eqnarray}
It is straightforward to check that under the above  gauge transformations, the first-order Lagrangian ($L_f$) 
remains invariant ($i.e., \; \delta L_f = 0 $).

\section {Off-shell nilpotent (anti-)BRST symmetry transformations and conserved charges}

The gauge-fixed and (anti-)BRST invariant Lagrangian for the Christ-Lee model can be written as \cite{kul}
\begin{eqnarray}
L = \dot r \,p_r + \dot \theta\, p_\theta - \frac{1}{2}\, p^2_r - \frac{1}{2 r^2}\, p^2_\theta - z \,p_\theta - V(r)
+  \frac{1}{2}\,b^2 +  b (\dot z + \theta)  - i\,\dot {\bar C}\, \dot C + i \, \bar C \, C,
\end{eqnarray}
where the Nakanishi-Lautrup auxiliary variable $b$ has been used to linearize the gauge-fixing term 
$-\frac{1}{2}(\dot z + \theta)^2$ as $\frac{b^2}{2} + b (\dot z + \theta)$. 
The anticommuting Faddeev-Popov ghost $(C)$ and anti-ghost $(\bar C)$ have ghost numbers $+1$ and $-1$, respectively, whereas
the remaining variables carry zero ghost number.
The above Lagrangian respects the off-shell nilpotent ($s^2_{(a)b} = 0 $) and absolutely 
anticommuting ($s_b\, s_{ab} + s_{ab}\, s_b = 0$) (anti-)BRST transformations ($s_{(a)b}$). These transformations are 
\begin{eqnarray}
&& s_b z = \dot C, \qquad s_b \theta = C, \qquad s_b \bar C = i\,b, \qquad s_b[r, \,p_r,\, p_\theta,\, b, \,C] = 0, \nonumber\\
&& s_{ab} z = \dot {\bar C}, \qquad s_{ab} \theta = \bar C, \qquad s_{ab}\, C = - i b, \qquad s_{ab}[r, \,p_r,\, p_\theta,b, \, \bar C] = 0.
\end{eqnarray}
It can be explicitly checked that under the above off-shell nilpotent symmetry transformations, the  Lagrangian $L$ remains  quasi-invariant. 
To be more precise, $L$ transforms to a total time derivative under the (anti-)BRST transformations as follows: 
\begin{eqnarray}
s_b\, L = \frac{d}{dt}\,\big[b \; \dot C\big], \qquad s_{ab}\, L = \frac{d}{dt}\,\big[b \; \dot {\bar C}\big].
\end{eqnarray}
As a result, both BRST and anti-BRST transformations leave the action integral ($S = \int dt L$) invariant ($i.e.,\;s_{(a)b}\, S = 0$).

According to Noether's theorem, the invariance of the action under the (anti-)BRST transformations leads to 
the following conserved charges (${\dot Q}_{(a)b} = 0$), namely
\begin{eqnarray}
Q_b &=& b\, \dot C + p_\theta \, C \equiv b\, \dot C - \dot b \, C, \nonumber\\
Q_{ab} &=& b\, \dot{\bar  C} + p_\theta \, \bar C \equiv b\, \dot{\bar  C} - \dot b \, \bar C.  
\end{eqnarray}
where on the r.h.s.,  we have used the equation of motion $p_\theta = - \,\dot b$.
These charges are nilpotent of order two ($i.e.,\; Q^2_b = 0,\, Q^2_{ab} = 0$) and anticommuting 
($i.e.,\; Q_b\,Q_{ab} + Q_{ab}\, Q_b = 0$)  in nature. The conservation laws for the (anti-)BRST charges can 
be proven by using the following  Euler-Lagrange equations of motion: 
\begin{eqnarray}
&&b = - (\dot z + \theta), \quad  \dot b + p_\theta = 0, \quad b = \dot p_\theta, \quad \dot p_r - \frac{p^2_\theta}{r^3} + V'(r) = 0, \nonumber\\
&& \dot r = p_r, \quad \dot \theta - z - \frac{p_\theta}{r^2} = 0, \quad \ddot C + C = 0, \qquad \ddot {\bar C} +  \bar C = 0.
\end{eqnarray}
The above equations of motion have been derived from Lagrangian (7).

It turns out that the (anti-)BRST charges are the generators of the (anti-)BRST symmetry transformations, 
respectively. As one can readily check, the following relations are true, namely
\begin{eqnarray}
s_b \Psi = - i \big[ \Psi,\; Q_b \big]_\pm,  \qquad s_{ab} \Psi = - i \big[\Psi, \;Q_{ab}\big]_\pm,
\end{eqnarray}
where $\Psi$ is any generic variable  present in the Lagrangian (7). The  $(\pm)$  signs as the subscript on the square brackets 
correspond to (anti)commutator depending on the generic variable $\Psi$ being (fermionic) bosonic in nature.

\section{(Anti-)co-BRST Symmetries and their generators}
We note that, in addition to the (anti-)BRST symmetries, the Lagrangian (7) also respects the following off-shell nilpotent
($i.e.,\; s^2_{(a)d} = 0$) and absolutely anticommuting ($i.e.,\; s_d\, s_{ad} + s_d \, s_{ad} = 0$) (anti-)co-BRST 
(or (anti-)dual-BRST] symmetries  ($s_{(a)d})$:
\begin{eqnarray}
&& s_d\, z = \bar C, \qquad s_d\, \theta = - \dot{\bar C}, \qquad s_d \,C =  i\, p_\theta, \qquad s_d\,[r,\; p_r,\; p_\theta,\; b,\; \bar C],\nonumber\\
&& s_{ad}\, z = C, \qquad s_{ad}\, \theta = - \dot{C}, \qquad s_{ad} \, \bar C = - i\, p_\theta, \qquad s_{ad}\,[r,\; p_r,\; p_\theta,\; b,\; C].
\end{eqnarray}
Under the above nilpotent symmetry transformations, the Lagrangian $L$ remains quasi-invariant, as one can check:  
\begin{eqnarray}
s_d\, L = - \frac{d}{dt}\big(p_\theta\,\dot {\bar C}\big), \qquad s_{ad}\, L = - \frac{d}{dt}\big(p_\theta\, \dot C).
\end{eqnarray}
Thus, the action  remains invariant ($i.e., \; s_{(a)d} S = 0$) under the application of (anti-)co-BRST transformations. 
It is worthwhile to mention here that the total gauge fixing-term $\frac{b^2}{2} + b \,(\dot z + \theta)$ 
remains invariant under the off-shell nilpotent (anti-)co-BRST transformations.

The invariance of the action under the continuous (anti-)co-BRST transformations leads to the following conserved (anti-)co-BRST charges 
$Q_{(a)d}$:  
\begin{eqnarray}
Q_d &=& b\, \bar C - p_\theta \, \dot {\bar C} \equiv b\, \bar C + \dot b \, \dot {\bar C}, \nonumber\\
Q_{ad} &=& b\, C - p_\theta \, \dot C \equiv b\, C + \dot b \, \dot C.  
\end{eqnarray}
The conservation laws ($\dot Q_{(a)d} =0$) of the (anti-)co BRST charges can be proven by exploiting the Euler-Lagrange equations of motion (11). 
It turns out that these charges are the generator of the (anti-)dual-BRST symmetries as one can check that the following relations are true:
\begin{eqnarray}
s_d \Psi = - i [\Phi,\; Q_d]_{\pm}, \qquad s_{ad} \Phi = - i [\Psi,\; Q_{ad}]_{\pm}, 
\end{eqnarray}
here $\Psi$ represents the generic variable present in the model. The ($\pm$) signs as the subscript on the square brackets have same meaning 
as mentioned in our previous section. We note that the (anti-)co-BRST charges are also nilpotent  ($Q^2_d = Q^2_{ab} = 0$) and anticommuting  
($Q_d\,Q_{ad} + Q_{ad}\, Q_d$ = 0) in nature.

\section{Bosonic symmetry and conserved charge}
Besides the above {\it four} fermionic symmetries as discussed in our earlier sections, we also have a {\it unique} 
bosonic symmetry present in the model. 
The bosonic symmetry $(s_\omega)$ is defined in terms of the fermionic symmetries  as given below
\begin{eqnarray}
s_\omega = \{s_b,\, s_d\} = - \{s_{ab},\, s_{ad}\}.
\end{eqnarray}
It is worthwhile to point out that rest of the anticommutators amongst the fermionic symmetries are explicitly zero. 
For the sake of completeness, these anticommutators  are as follows
\begin{eqnarray}
&&\big\{s_b,\, s_{ab} \big\} = 0, \qquad  \quad\big\{s_b,\, s_{ad} \big\}= 0,  \nonumber\\
&& \big\{s_d, \; s_{ab} \big\} = 0, \qquad \quad  \big\{s_d,\, s_{ad}\big\} = 0.
\end{eqnarray}
As a consequence, the above vanishing anticommutators do not define the symmetry.  
The bosonic symmetry transformations for all the variables are as follows:
\begin{eqnarray}
&& s_\omega\, z = + i\,\big(b + \dot p_\theta\big), \qquad s_\omega\, \theta = - i\,\big(\dot b - p_\theta\big), \qquad
s_\omega [r, \;p_r, \; p_\theta, \; b, \; C, \;  \bar C] = 0.
\end{eqnarray}
Under the above bosonic symmetry, $L$ transforms to a total time derivative as 
\begin{eqnarray}
s_\omega \, L = \frac{d}{dt}\big[i \,\big(b\, {\dot p}_\theta - \dot b\, p_\theta \big) \big].
\end{eqnarray}
Thus, the action integral remains invariant. According to Noether's  theorem, the above continuous 
bosonic symmetry leads to the derivation of the following conserved charge: 
\begin{eqnarray}
Q_\omega =  i\, \big(b^2 + p^2_\theta)  \;\equiv\;  i\, \big(b\, {\dot p}_\theta - \dot b \,p_\theta). 
\end{eqnarray}
The above charge is the generator of bosonic symmetry $s_\omega$ as one can check the following transformations $s_\omega \Psi = - i [\Psi,\, Q_\omega]$
is true for any generic variable $\Psi$.

\section{Ghost scale and discrete symmetries}

It is straightforward to check that the following ghost-scale symmetry transformations:
\begin{eqnarray}
&& C \to e^{+1 \cdot \lambda}\,C, \quad \quad \bar C \to e^{- 1 \cdot \lambda}\, \bar C,  \nonumber\\
&&(r, \theta, z, p_r, p_\theta, b) \to e^{0\cdot \lambda} \,(r, \theta, z, p_r, p_\theta, b),
\end{eqnarray}
leave the Lagrangian (7) invariant. Here $\lambda$ is a time independent (global) scale parameter. 
The numerals in the exponential represent the ghost number of 
the corresponding variables. In particular, the ghost numbers of $(C) \bar C $ are $(+ 1) - 1$, respectively  
whereas remaining variables have zero ghost number.   
Under the infinitesimal version of the above symmetry
(with $\lambda = 1$)
\begin{eqnarray}
s_g\, C = + C \qquad s_g\,\bar C = - \bar C, \qquad s_g\,[r, \theta, z, p_r, p_\theta, b] = 0,
\end{eqnarray}
the Lagrangian remains invariant ($i.e., \; s_g L = 0$) and thus we obtain conserved ($\dot Q_g = 0$) ghost charge as follows:
\begin{eqnarray}
Q_g = i \big(\bar C \, \dot C - \dot {\bar C}\, C\big).
\end{eqnarray}
It is straightforward to check that $Q_g$ is the generator of the above infinitesimal ghost scale symmetry transformations.  
Besides the above continuous ghost scale symmetries, the ghost part of the Lagrangian (7) has the following
discrete symmetries: $C \to \pm  \bar C$ and $\bar C \to \mp C$. The latter symmetries play a decisive role in obtaining the 
anti-BRST and anti-co-BRST symmetries from the BRST and co-BRST symmetries, respectively. 
Further, we also note that under the discrete symmetry transformations, the ghost charge remains invariant.

\section {Extended BRST algebra and cohomological aspects} 
Exploiting the operator form of the six independent continuous symmetries listed in equations (8), (13), (19) and (23), we obtain the following 
extended BRST algebra:
\begin{eqnarray}
&&  s^2_{(a)b} = 0, \qquad  s^2_{(a)d} = 0, \qquad  \big\{s_b,\, s_{ab} \big\} = 0, \nonumber\\
&&  \big\{s_d,\, s_{ad} \big\} = 0,\qquad \big\{s_b,\, s_{ad} \big\} = 0, \qquad \big\{s_{ab},\, s_d \big\} = 0, \nonumber\\
&& \big[s_g,\, s_b \big] = s_b, \qquad \big[s_g, s_{ab} \big] = -s_{ab}, \qquad  \big[s_g, s_d \big] = -s_d, \nonumber\\
&& \big[s_g, s_{ad}\big] = s_{ad},  \qquad \big\{s_b,\, s_d \big\} =  - \big\{s_{ab},\, s_{ad} \big\} = s_\omega, \nonumber\\
&& \big[s_\omega, \,s_r] = 0,  \;\;\qquad r = b, ab, d, ad, g.
\end{eqnarray}
It can also be checked that the above similar type of algebra is satisfied by the conserved charges as given below: 
\begin{eqnarray}
&& Q^2_{(a)b} = 0, \qquad  Q^2_{(a)d} = 0, \qquad  \big\{Q_b, \;Q_{ab}\big\} = 0, \nonumber\\ 
&& \big\{Q_d, \,Q_{ad}\big\} = 0, \qquad \big\{Q_b,\, Q_{ad}\big\} = 0, \qquad \big\{Q_d, \,Q_{ab} \big\} = 0, \nonumber\\
&& \big[Q_g, \,Q_b\big] = - i \,Q_b, \qquad \big[Q_g,\, Q_{ad}\big] = - i\, Q_{ad}, \nonumber\\
&& \big[Q_g, \,Q_{ab}\big] = + i\, Q_{ab}, \qquad \big[Q_g, \,Q_d\big] = + i\,Q_d, \nonumber\\
&& \big\{Q_b,\, Q_d\big\} = -  \big\{Q_{ab},\, Q_{ad}\big\}  = Q_\omega, \nonumber\\
&& \big[Q_\omega,\; Q_r\big] = 0,
\end{eqnarray}
where, in proving the above algebra amongst the conserved charges, we have used the following basic (anti)commutators:  
\begin{eqnarray}
&& \big[r,\; p_r\big] = i, \qquad \big[\theta,\; p_\theta\big] = i, \qquad \big[z,\; b\big] = i,\nonumber\\
&& \big\{C, \; \dot {\bar C}\big\} = 1, \qquad \big\{\bar C, \; \dot C\big\} = - 1,
\end{eqnarray}
and the remaining (anti)commutators are turn out to be zero. We point out that the  algebra (26) can also be proven in a simple 
and straightforward manner by exploiting the definition of generator and the continuous symmetries (and corresponding charges).  
For instance, the  nilpotency and anticommutativity properties  of the fermionic charges can also be proved as follows:
\begin{eqnarray}
s_b Q_b &=& -\, i \big\{Q_b, \, Q_b \big\} = 0 \;\Rightarrow\; Q_b^2 = 0, \nonumber\\
s_{ab} Q_{ab} &=& - \, i \big\{Q_{ab}, \, Q_{ab}\big\}  = 0 \;\Rightarrow\; Q_{ab}^2 = 0, \nonumber\\
s_b Q_{ab} &=& - \, i \big\{Q_{ab}, \, Q_b \big\} = 0 \Rightarrow Q_b\, Q_{ab} + Q_{ab}\, Q_b = 0, \nonumber\\
s_{ab} Q_b &=& - \, i \big\{Q_b, \, Q_{ab} \big\} = 0 \Rightarrow Q_b \,Q_{ab} + Q_{ab} \,Q_b = 0. \quad
\end{eqnarray}
Similarly, one can also compute other relations that appear in eq. (26).

It is to be noted that the algebra given in (25) and (26) are reminiscent of the Hodge algebra obeyed by de Rham cohomological 
operators ($d, \delta, \Delta $) of differential geometry. The later algebra is  \cite{NO, EGH, MN, NISH, dght}
\begin{eqnarray}
&&d^2 = 0, \qquad \delta^2 = 0, \qquad \big\{d, \delta\big\} = \Delta  = d\, \delta + \delta \, d, \nonumber\\ 
&& \big[\Delta, \, d\big] = 0, \qquad \big[\Delta, \, \delta \big] = 0.
\end{eqnarray}
where $(\delta) d$ are the (co-)exterior derivatives and $\Delta$ is the Laplacian operator. 
A close look at the above algebra and the algebra obeyed by the continuous symmetry transformations (and corresponding charges) shows
that there is a one-to-two mapping between the cohomological operators and the symmetry transformations (their corresponding conserved charges)
\begin{eqnarray}
&& (s_b, s_{ad}) \to d, \qquad  (s_{ab}, s_d) \to \delta, \nonumber\\
&& s_\omega = \big\{s_b, \,s_d\big\} = - \big\{s_{ab}, \,s_{ad}\big\} \to \Delta,\nonumber\\
&& \nonumber\\
&& (Q_b, Q_{ad}) \to d, \qquad  (Q_{ab}, Q_d) \to \delta, \nonumber\\ 
&&Q_\omega = \big\{Q_b, \,Q_d\big\} = - \big\{Q_{ab}, \,Q_{ad}\big\} \to \Delta.
\end{eqnarray} 
As a consequence,  the continuous symmetry transformations (and their corresponding charges) provide the physical realizations of the abstract
mathematical de Rham cohomological operators. Thus, the present Christ-Lee  model provides  a  model for Hodge theory.

As far as the mathematical properties of the cohomological operators are concerned, we note that the exterior derivative  $d$,
when acts on any given $n$-form $f_n$ of degree $n$, increases the degree of the form by one unit ($i.e., \;d f_n \sim f_{n+1}$) whereas the co-exterior
derivative $\delta$ does its opposite when it operates on $f_n$ ($i.e.,\; \delta  f_n \sim f_{n-1}$). 
Further, the Laplacian operator $\Delta$ does not affect the degree of the form when it operates on a given form  
($i.e., \; \Delta  f_n \sim f_n$).  These sacrosanct properties can be captured in terms of ghost number.    If we look carefully 
on symmetry transformations (8), (13) and (17), we note that the sets  $(s_b, \, s_{ad})$ and $(s_{ab}, \, s_d)$ increase and  
decrease the ghost numbers by one, respectively when they operate on any generic variable while $s_\omega$ does not change 
the ghost number.  We can also capture  these properties in terms of the conserved charges. For this purpose, 
we define a state $|\psi\rangle_n$ (in the total quantum Hilbert space of states) as follows:    
\begin{eqnarray}
i\, Q_g | \psi \rangle_n = n |\psi \rangle_n,
\end{eqnarray}
where $n$ defines the ghost number of the state $|\psi\rangle_n$ as the eigenvalue of the operator $i Q_g.$  Using the algebra (26) 
and validity of  the eq. (31), we yield the following relations:
\begin{eqnarray}
i\, Q_g\, Q_b |\psi \rangle_n &=& (n + 1) Q_b\, |\psi \rangle_n,\nonumber\\
i\, Q_g \,Q_{ad} |\psi \rangle_n  &=& (n + 1)\, Q_{ad} |\psi \rangle_n, \nonumber\\
i\, Q_g \,Q_d |\psi \rangle_n &=& (n - 1)\, Q_d |\psi \rangle_n, \nonumber\\
i \,Q_g \,Q_{ab} |\psi \rangle_n &=&(n - 1)\, Q_{ab} |\psi \rangle_n, \nonumber\\
i \,Q_g\, Q_\omega |\psi \rangle_n &=& n \,Q_\omega |\psi \rangle_n. 
\end{eqnarray}
It evident that the states $Q_b |\psi\rangle_n$,   $Q_d |\psi\rangle_n$ and 
$Q_\omega |\psi\rangle_n$ are endowed with the ghost numbers equal to $(n+ 1)$, $(n- 1)$ and $n$, respectively. 
Similarly, the states $Q_{ad} |\psi\rangle_n$,   $Q_{ab} |\psi\rangle_n$ and $Q_\omega |\psi\rangle_n$
have ghost numbers $(n+ 1)$, $(n- 1)$ and $n$, respectively. Thus, the following sets 
$(Q_b, \, Q_d, Q_\omega)$ and $(Q_{ad}, \, Q_{ab}, - Q_\omega)$, as the conserved charges present in our model, 
provide the physical realizations of the de Rham cohomological operators  $(d, \,\delta,\, \Delta)$ of differential geometry.

We wrap up this section with the remarks that, in the realm of BRST quantization,  the physicality criteria $Q_{(a)b}|phys\rangle = 0$ 
lead to the following requirements:  $b|phys\rangle = 0$ and $\dot b|phys\rangle = 0$. Due to the validity of the Euler-Lagrange equations of
motion (11), these requirements imply that the operator form of the first-class constrains $p_z \approx 0$ and $p_\theta \approx 0$
present in the original singular Lagrangian annihilate the physical state ($|phys\rangle$). In addition, 
it is interesting to note that the physicality conditions $Q_{(a)d}|phys\rangle = 0$ also produce the same requirements. 
Thus, the physicality criteria $Q_{(a)b}|phys \rangle = 0$ and $Q_{(a)d}|phys \rangle = 0$ are consistent with the Dirac's 
quantization of constrained system.

\section{Conclusions}

In our present investigation, we have discussed about the Christ-Lee model within the framework of BRST formalism. We have shown that, 
in addition to the nilpotent BRST and anti-BRST symmetries, there also exist {\it two} more nilpotent symmetries, namely; co-BRST and anti-co-BRST
transformations (at quantum level). The non-vanishing anticommutators amongst the fermionic (anti-)BRST and 
(anti-)co-BRST symmetry transformations yield a unique bosonic symmetry.  
Furthermore, a continuous ghost-scale symmetry also exists for the present model. Thus, within the framework of BRST formalism, we have
{\it six} continuous symmetry transformations. These continuous symmetries, according to Noether's theorem, lead to the derivation 
of corresponding conserved charges. By exploiting the definition of a generator, we have also proven the nilpotency and anticommutativity properties  of 
the fermionic charges in simpler way (cf. (28)).

The algebra satisfied by the operator form of the continuous symmetry transformations ($s_{(a)b}, \, s_{(a)d}$ and $s_\omega$) 
is the replicate of the algebra obeyed by de  Rham cohomological operators ($i.e., \;d$, $\delta$ and $\Delta$)
 of differential geometry (cf. (25) and (29)). The similar algebra is also satisfied by the conserved charges 
($Q_{(a)b}, \, Q_{(a)d}$ and $Q_\omega$) (cf. (26)). On the basis of the above, finally, we have shown
that the de Rham cohomological operators find their physical meanings in terms of the continuous symmetry transformations (and corresponding 
conserved charges). In fact, we have two-to-one mapping between the symmetry transformations (and their corresponding generators) and the 
de Rham cohomological operators.  Thus, the Christ-Lee model provides us a simple toy model for the Hodge theory within the framework 
of BRST formalism.

It would be a nice endeavor to derive the continuous transformations ($s_{(a)b}, \, s_{(a)d}$) by exploiting the
celebrated supervariable approach where these transformations get their geometrical interpretation in terms
of the Grassmannian derivatives. Furthermore, the invariance of Lagrangian and the nilpotency of charges can also be captured
in terms of the supervariables and Grassmannian derivatives. It would also be interesting to derive the basic canonical brackets
by exploiting the use of nilpotent charges within the framework of BRST formalism where one does not need any recourse of the 
canonical momenta. These are the some important issues for the present Christ-Lee model which we would like 
to address in our future publications elsewhere.

In our earlier works \cite{rpm2,rpm3,rpm4,rpm5,rpm6,rpm7,rpm8}, we have shown that any arbitrary 
Abelian $p$-form ($p =1,2,3$) gauge theory in $D =2p$ dimensions of spacetime within the framework of BRST formalism and 
the ${\cal N} = 2$ SUSY quantum mechanical models turn out to be the tractable model for Hodge theory. It would be interesting 
to implement this idea in the case of a free particle system on a toric geometry \cite{rk},  
supersymmetric Yang-Mills \cite{mir7,mir11,mir12,mir13}. Furthermore, the derivation of proper (anti-)BRST symmetries with the help of
superfield formalism  would be a nice piece of work in the context of deformed super-Yang-Mills, 
supersymmetric Chern-Simons, ABJM and  BLG theories \cite{mir13,mir14,mir15}.

\section*{Acknowledgments:} RK would like to thank UGC, Government of India, New Delhi, for financial support under the PDFSS scheme.

\end{document}